\documentclass{aip-cp}
\usepackage[numbers]{natbib}
\usepackage{rotating}
\usepackage{graphicx}

\begin{document}

% Title portion
\title{Point Source Detection and Flux Determination with PGWave}

\author[aff1,aff2]{Giacomo Principe\corref{cor1}}
\author[aff1,aff2]{Dmitry Malyshev}

\affil[aff1]{on behalf of Fermi Large Area Telescope Collaboration}
\affil[aff2]{Erlangen Centre for Astroparticle Physics ECAP, D-91058 Erlangen, Germany }

\corresp[cor1]{Corresponding author: giacomo.principe@fau.de}

\maketitle

\begin{abstract}
One of the largest uncertainties in the Point Source (PS) studies, at Fermi-LAT energies \cite{atwood}, is the uncertainty in the diffuse background. In general there are two approaches for PS analysis: background-dependent methods, that include modeling of the diffuse background, and background-independent methods. In this work we study PGWave \cite{tosti}, which is one of the background-independent methods, based on wavelet filtering to find significant clusters of gamma rays. PGWave is already used in the Fermi-LAT catalog pipeline for finding candidate sources. We test PGWave, not only for source detection, but especially to estimate the flux without the need of a background model. We use Monte Carlo (MC) simulation to study the accuracy of PS detection and estimation of the flux. We present preliminary results of these MC studies.
\end{abstract}

% Head 1
\section{INTRODUCTION}
In the X-ray and gamma-ray astronomy one of the most important problems is to provide an accurate and precise diffuse model to detect PS and to estimate their spectra. If the diffuse model does not correctly reproduce the background of the sky, it can introduce large systematic errors in the detection of the PS and in the flux determination. This is the motivation to find a background-independent method that is not influenced by the systematic error due to the precision of the diffuse model.

\section{PGWave: a Wavelet Transform Method}\label{wav_met}
PGWave is a method, based on Wavelet Transforms (WTs) \cite{damiani}, that can be used to find PS without the need of a background model.

The WT of a 2-dim image $f(x,y)$ is defined as:
\begin{equation}
w(x,y,a) = \iint g (\frac{x-x'}{a} , \frac{y-y'}{a}) f(x',y') dx' dy' \; ,
\end{equation}
where $g(x/a,y/a)$ is the generating wavelet, $x$ and $y$ are the pixel coordinates, and $a$ is the scale parameter.

\begin{figure}[h]
  \centerline{\includegraphics[width=150pt]{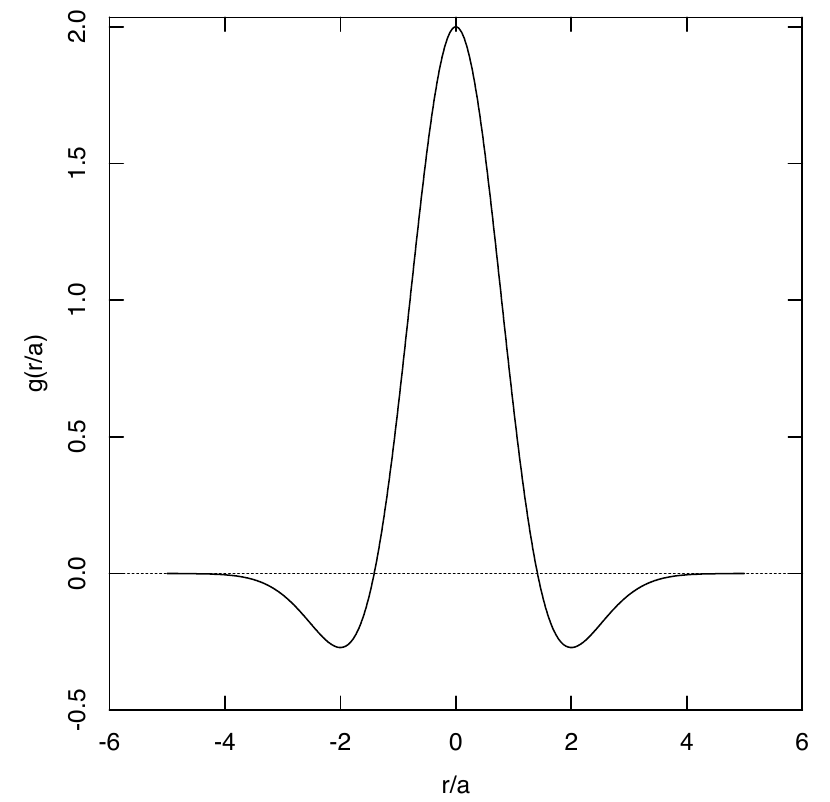}}
  \caption{Mexican hat generating wavelet $g(r/a)$.}
  \label{mex_hat}
\end{figure}

PGWave uses the 2-dim “Mexican Hat” wavelet (Figure \ref{mex_hat}):
\begin{equation}
g (\frac{x}{a} , \frac{y}{a}) = g(\frac{r}{a}) = (2-\frac{r^{2}}{a^{2}}) \, e^{-r^{2}/2a^{2}}  \:  (r^{2} = x^{2} + y^{2}) \; .
\end{equation}

This chosen $g$ ensures that the WT of a function $f(x, y) = c_{1} + c_{2} x + c_{3} y$ (a tilted plane) is zero. Therefore, the WT will be zero for both a constant or uniform gradient local background. While the WT is not sensitive to gradients in the data with this choice of $g$, it is sensitive to second derivatives of $f (x, y)$, e.g. local maximum or minimum, making it suitable for detecting PS in an image.

The peak of the WT for a source with Gaussian shape ($N_{src}$ total counts and width $\sigma_{src}$) is  \cite{damiani}:
\begin{equation} 
w_{peak} (a) = \frac{2 N_{src}}{(1+\sigma_{src}^{2}/a^{2})^{2}} \; .
\end{equation}
Such dependence shows a linear correlation between the WT peak and the total number of photons detected from a source.

\section{Monte Carlo Simulated Data}\label{mc_data}
The data used to study PGWave are created using MC simulation of the sky in the energy range from 1 GeV to 10 GeV.
The MC simulations are performed with \textit{gtobssim}, a Fermi tool used to simulate point sources and diffuse emission for Fermi observations, with a specific spectral shape, for a selected region of the sky. 
\footnote{We thank Mattia di Mauro (SLAC) for providing us the MC data.}

The simulation includes:
\begin{itemize}
\item galactic diffuse emission,
\item isotropic background,
\item PS.
\end{itemize}

\begin{figure}[h]
  \centerline{\includegraphics[width=230pt]{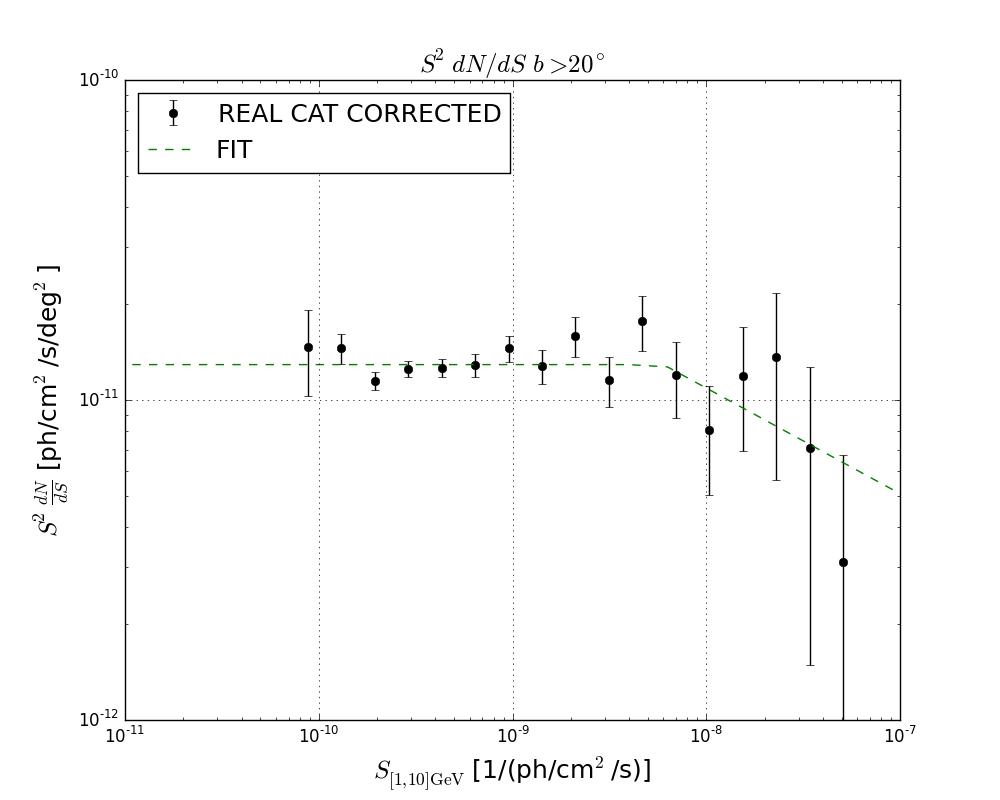}}
  \caption{$dN/dS$ distribution of 3FGL PS. The green dashed line is the input PS distribution in the Monte Carlo simulation.}
 \label{dnds}
\end{figure}

The fluxes of the PS follow the $LogN-LogS$ function given by the green dashed line in Figure \ref{dnds}, derived from 3FGL catalog \cite{acero}. The spectral energy distribution of the sources is a Power law with spectral index randomly chosen from Gaussian distribution with $\mu  = 2.30$ and $\sigma = 0.40$  motivated by the distribution of extra-galactic sources in the 3FGL catalog.

The main parameters used to create the MC data are summarized in Table \ref{par_mc}.

% Table
\begin{table}[h]
\label{par_mc}
\tabcolsep7pt\begin{tabular}{cc}
\hline
Energy range & 1 GeV - 10 GeV \\
\hline
Interval of time & 92 months \\
\hline 
IRF & P8R2$_{-}$SOURCE$_{-}$V6\\
\hline
\end{tabular}
\caption{Parameters used to create the MC data.}
\end{table}

\begin{figure}[h]
 \centerline{\includegraphics[width=300pt]{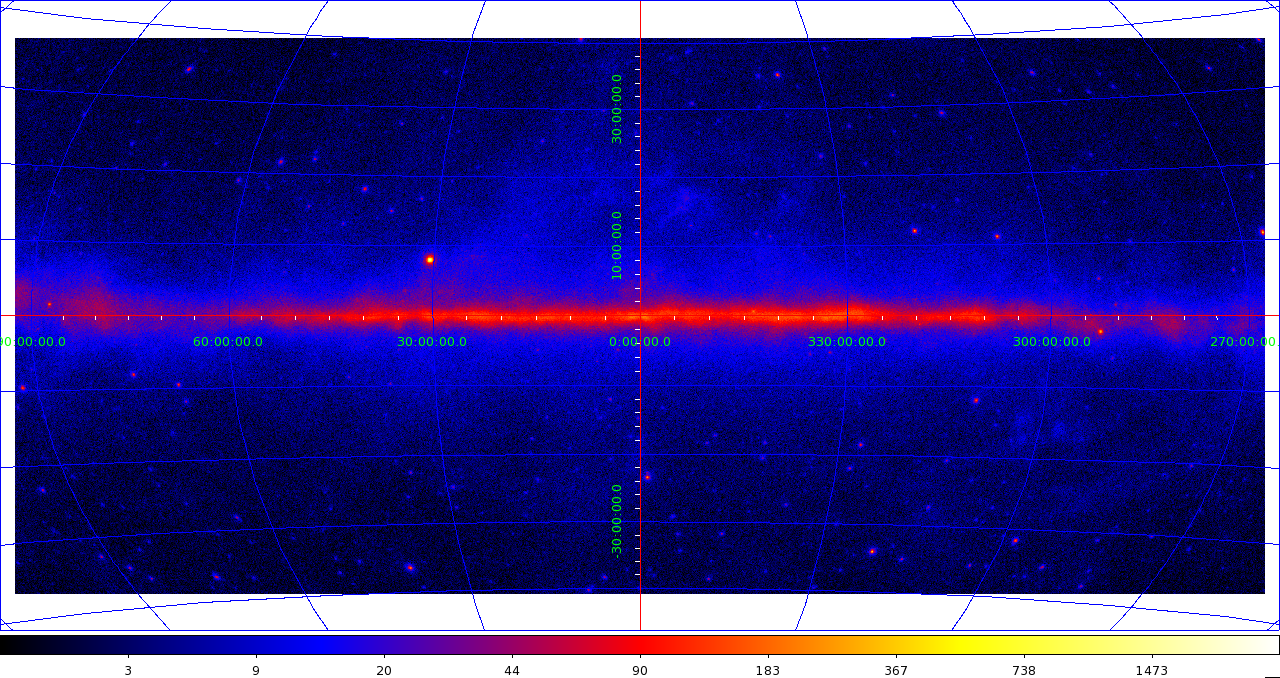}}
  \caption{Counts map of the MC data in the region of the sky: $-90^{\circ}<LON<90^{\circ}$, $-40^{\circ}<LAT<40^{\circ}$.}
\label{mc_plot}
\end{figure}

\section{PGWave: Analysis}\label{pgwave_analysis}
PGWave, as described before, is a WT method that works on counts map. We pass the MC data as input in PGWave and it gives a list of PS candidates that contains, in particular, position, significance and WT peak values.

For the analysis with PGWave we apply some selections to the data:
\begin{enumerate}
\item we restrict the analysis in the area: $-90^{\circ}<LON<90^{\circ}$, $-40^{\circ}<LAT<40^{\circ}$,
\item we mask the Galactic Plane ($-5^{\circ}<LAT<5^{\circ}$),
\item we choose MC sources with a Flux $> 10^{-10} \mathrm{ph\, cm^{-2} s^{-1}} $.
\end{enumerate}

The parameters used for PGWave analysis are summarized in Table \ref{tab:pg_an}. 

% Table
\begin{table}[h]

\label{tab:pg_an}
\tabcolsep7pt\begin{tabular}{cc}
\hline
Pixel dimension & $0.1^{\circ}$ \\
\hline
MH Wavelet Transform scale & $0.3^{\circ}$\\
\hline 
Statistical confidence & 3 $\sigma$\\
\hline
Minimum number of connected pixels & 5\\
\hline
\end{tabular}
\caption{Parameters used in the PGWave analysis.}
\end{table}

\subsection{PGWave: Point Source Detection}
We compare the results with the MC data using a method that associates the PGWave seeds with the input sources. The algorithm for the association is based on a positional coincidence, with a tolerance radius of $0.56^{\circ}$ (similar to PSF at 2 GeV), and on a flux ordering. The flux ordering algorithm associates the seeds with a large WT peak value with the bright MC input sources, this allows a better association in particular in the cases where inside the tolerance radius more than one source are present.
 
In the MC data there are 1230 sources, with a flux above $> 10^{-10} \mathrm{ph\, cm^{-2} s^{-1}} $. PGWave finds 808 seeds, of which 720 are associated with MC sources using our association method with a tolerance radius of $0.56^{\circ}$.

% INSERT FIGURE MC data plot
\begin{figure}[h]
 \includegraphics[width=270pt]{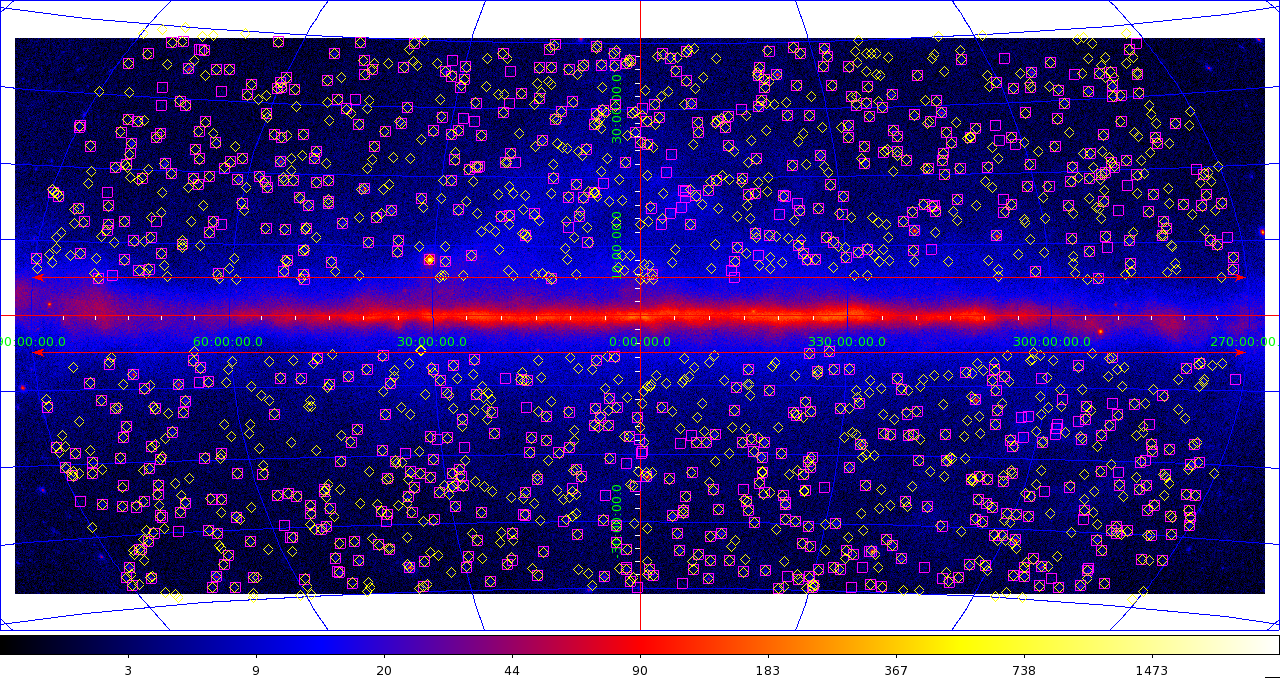}
 \includegraphics[width=270pt]{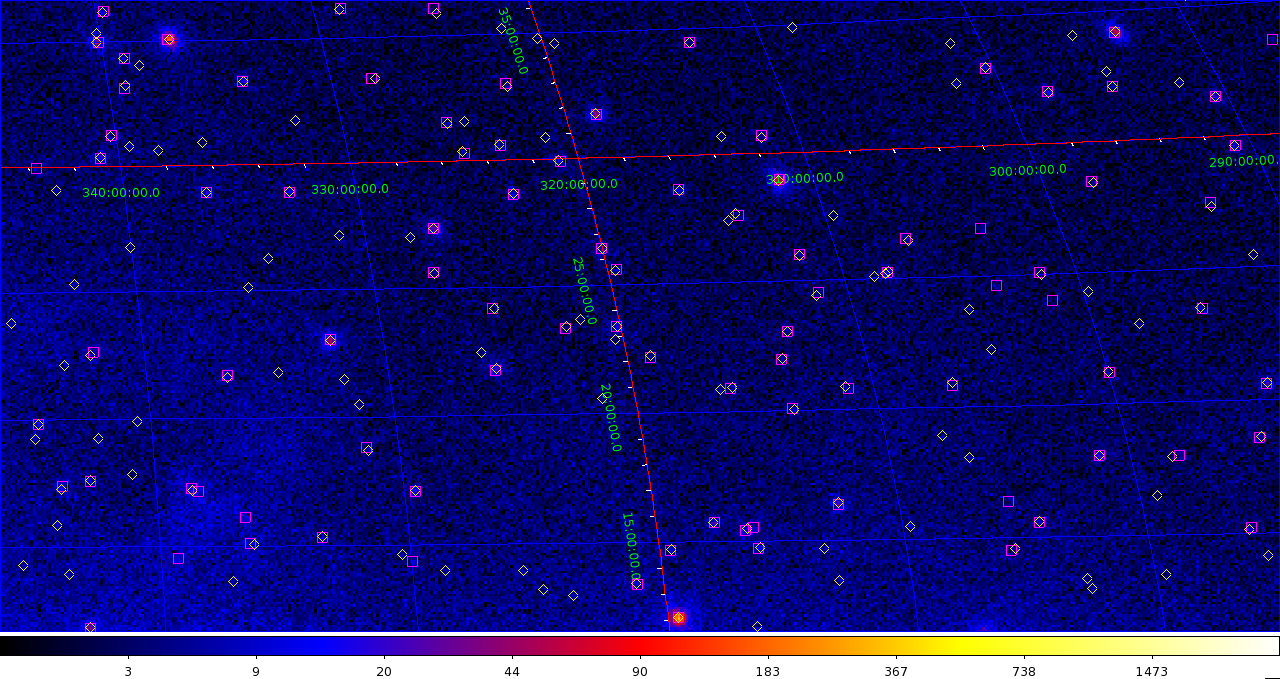}
  \caption{Counts maps of the MC data with MC sources, represented here with yellow diamonds,  and PGWave seeds, represented with magenta diamonds. Left: the counts map of all the region. Right: a zoom in the region of the dimensions of $40 \times 60$ degrees centered in the position of LAT$=25^{\circ}$ and LON$=320^{\circ}$ .}
 \label{ps_map}
\end{figure}

\subsection{PGWave: Flux Determination} 
\begin{figure}[h]
  \includegraphics[width=300pt]{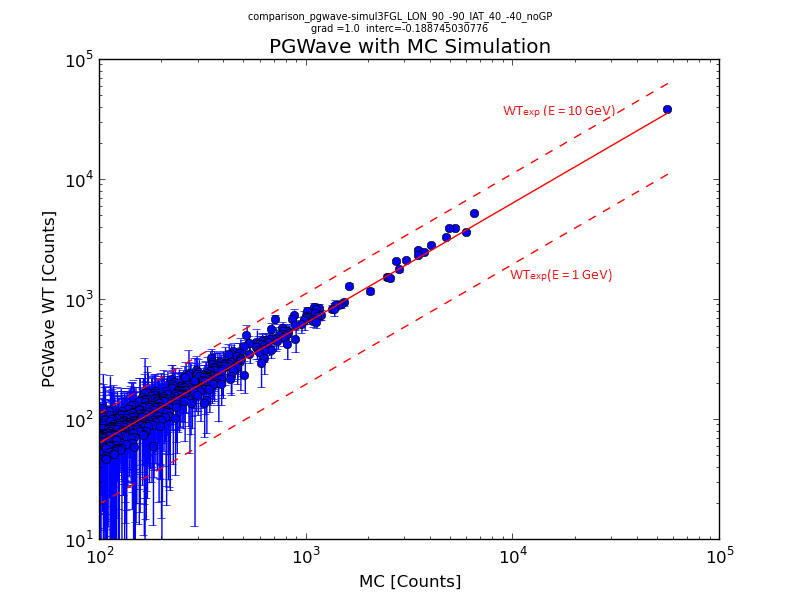} \label{flux_det}
  \includegraphics[width=300pt]{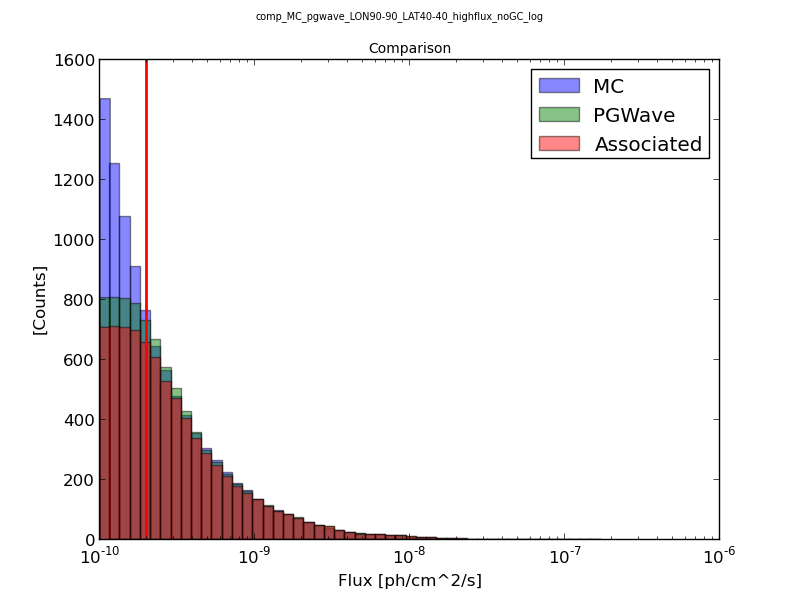} 
  \caption{Left: Plot of the associated PS. X-axis: MC input counts, y-axis: the WT peak value from PGWave. The red line represents the best fit, the two dashed lines are the expected values of the WT peak for the energy of 1 GeV and 10 GeV. Right: Inverse cumulative distribution of the associated PS with the estimated flux. The red line represents the cut applied for the best fit. }
  
\end{figure}

For the estimation of the flux of the associated PS we use the WT peak value, because a linear correlation between them is expected (Equation(3)).
Figure \ref{flux_det} (left) shows the plot of the MC input flux vs the WT peak value of the associated PS. A linear correlations is observed. This allows the possibility to derive the flux of the sources using a fit and the WT peak values provided by PGWave. To determine the best fit, we select the associated PS with MC Counts $>$100.

The statistical uncertainty on the WT peak value is estimated by the total number of photons inside the circle with wavelet scale radius.

To derive the bracketing values we use the PSF at 1 GeV and at 10 GeV for the P8R2$_{-}$SOURCE$_{-}$V6 data in the WT peak (Equation (3)). All reconstructed PS fluxes fall within the bracketing values. 

One can see from Figure \ref{flux_det}(right) that PGWave finds more than $85 \%$ of the MC sources above $2 \times 10 ^{-10}  \mathrm{ph\, cm^{-2} s^{-1}}$.

\section{Conclusion}
\label{conclusion}
In this work we apply PGWave to MC simulated data, in the presence of diffuse emission background.
\begin{enumerate}
\item We confirm that PS detection works well at $\mid LAT \mid > 5^{\circ}$.
\item There is a very good correlation between the WT peak and the input MC flux which makes WT a very promising method to find a background independent estimate of the PS flux.
\end{enumerate}
In future, PGWave could be useful for detecting PS and estimating their flux without the need of a background model.

% Acknowledgement
\section{ACKNOWLEDGMENTS}
The Fermi-LAT Collaboration acknowledges support for LAT development, operation and data analysis from NASA and DOE (United States), CEA/Irfu and IN2P3/CNRS (France), ASI and INFN (Italy), MEXT, KEK, and JAXA (Japan), and the K.A. Wallenberg Foundation, the Swedish Research Council and the National Space Board (Sweden). Science analysis support in the operations phase from INAF (Italy) and CNES (France) is also gratefully acknowledged.

% References

\nocite{*}
\bibliographystyle{aipnum-cp}%
\bibliography{sample}%

\begin{thebibliography}{1}
\bibitem{atwood} Atwood, W. B. et al., \textit{The Large Area Telescope on the Fermi Gamma-ray Space Telescope Mission}, ApJ, 697, 1071 (2009).
\bibitem{tosti} Tosti G. et al., \textit{Gamma-ray sources detection using PGWave}, Fermi Coll. Meeting SLAC (2005)
\bibitem{damiani} Damiani F. et. al., \textit{A Method Based on Wavelet Transforms for Source Detection in Photon-Counting Detector Images}, ApJ 483, 350,(1997)
\bibitem{acero} Acero F. et al., \textit{Fermi Large Area Telescope Third Source}, arXiv:1501.02003 (2015)

 \end{thebibliography}

\end{document}